\documentclass[aps, prl, twocolumn, amsmath,amssymb,superscriptaddress]{revtex4-2}

\usepackage{hyperref}
\usepackage{graphicx}
\usepackage{dcolumn}
\usepackage{xcolor}
\usepackage{braket}
\usepackage{commath}
\usepackage{adjustbox}
\usepackage{comment} 
\usepackage{ulem}
\usepackage{tikz}
\usetikzlibrary{shapes}

\newcommand{\bs}[1]{{\boldsymbol{#1}}}
\newcommand{\bk}{\bs{k}}

\newcommand{\br}{\bs{r}}

\begin{document}

\title{Dynamical emergence of a Kosterlitz-Thouless transition in a disordered Bose gas following a quench}

\author{Thibault Scoquart}
\affiliation{Laboratoire Kastler Brossel, Sorbonne Universit\'e, CNRS, ENS-PSL University, Coll\`ege de France; 4 Place Jussieu, 75005 Paris, France}
\affiliation{Institut f\"ur Theorie der Kondensierten Materie, Karlsruhe Institute of Technology, 76128 Karlsruhe, Germany}
\author{Dominique Delande}
\affiliation{Laboratoire Kastler Brossel, Sorbonne Universit\'e, CNRS, ENS-PSL University, Coll\`ege de France; 4 Place Jussieu, 75005 Paris, France}
\author{Nicolas Cherroret}
\affiliation{Laboratoire Kastler Brossel, Sorbonne Universit\'e, CNRS, ENS-PSL University, Coll\`ege de France; 4 Place Jussieu, 75005 Paris, France}


\begin{abstract}
We study the dynamical evolution of a two-dimensional Bose gas after a disorder potential quench. Depending on the initial conditions, the system evolves either to a thermal or a superfluid state. 
Using extensive quasi-exact numerical simulations, we show that the two phases are separated by a Kosterlitz-Thouless transition. The thermalization time is shown to be longer in the superfluid phase,
but no critical slowing down is observed at the transition. The long-time phase diagram is well reproduced by a simple theoretical model.  
The spontaneous emergence of Kosterlitz-Thouless transitions following a quench is a generic phenomenon that should arise both in the context of non-equilibrium quantum gases and nonlinear, classical wave systems.
\end{abstract}

\maketitle
 
Following a quench, 
ergodic quantum systems 
experience a progressive loss of memory of their initial state. At long time a thermal equilibrium establishes, governed by a few conserved quantities \cite{Polkovnikov11, Gogolin2016}. This ubiquitous phenomenon arises, in particular,  in isolated systems, which act as a thermal bath for their subparts, as formalized by the Eigenstate Thermalization Hypothesis \cite{DAlessio2016, Deutsch2018, Ueda2020}. Despite the rather generic character of thermalization, it has been shown that its transient dynamics could be very rich. For certain far-from-equilibrium initial states, e.g., the post-quench correlations functions of a quantum gas can display a universal spatio-temporal scaling  
 \cite{Berges2008, Pineiro2015, Mikheev2019}. Near-integrable systems also exhibit a slow pre-thermal dynamics \cite{Berges2004, Trotzky2012, Abuzarli2022, Gring2012}, characterized by a generalized Gibbs ensemble \cite{Kollar11, Langen2015}. Theoretically, descriptions of the full evolution of a quantum system from its early- to late-time dynamics have been proposed in a few cases \cite{Regemortel2018, Mallayya2018, Mallayya2019, Buchhold2016} but this problem remains, in general, largely unexplored.

An important ingredient that may significantly impact the quench dynamics of an isolated system is spatial disorder. The role of disorder has been, for instance, addressed in the context of strongly interacting systems where the phenomenon of many-body localization prevents the emergence of a thermal state \cite{Abanin2019}. Initially described in one-dimensional geometries \cite{Gornyi2005, Basko2006}, many-body localization has also been recently touched upon in two dimensions \cite{Choi2016, Bertoli2018, Theveniaut2020, Pietracaprina2021}. In parallel, the ergodic regime has been explored in weakly interacting disordered Bose gases, where it was shown that at transient times the competition between disorder and interactions  can destroy localization effects \cite{Kopidakis2008, Pikovsky2008, Cherroret2014, Scoquart2020} or drive the gas to a pre-thermal state \cite{Scoquart2020b, Cherroret2021}.

Another central question in non-equilibrium physics is the possibility for coherent, condensate-like structures to spontaneously emerge in an isolated system after a quench, a phenomenon that depends on dimensionality and occurs for specific initial conditions 
 \cite{Connaughton2005, Berges2012, Cherroret2015, Chiocchetta2016, Chiocchetta2016b, Shukla2021, Haldar2021, Nazarenko2011}. 
In practice, the dynamical formation of condensates has been much studied in the context of nonlinear optics where an optical field, analogously to a degenerate Bose gas, obeys a time-dependent nonlinear Schr\"odinger equation. Such `dynamical' condensation has been explored in atomic vapors \cite{Santic2018, Sun2012} and in multimode fibers \cite{Wright2016, Krupa2017, Baudin2020}. 
In these works, the effective spatial dimension was 2, and signatures of condensation were observed either due to the presence of a spatial confinement \cite{Wright2016, Krupa2017, Baudin2020} or because the dynamics was probed at  finite time \cite{Santic2018, Sun2012}. In two-dimensional (2D) infinite space, however, no \emph{true} condensation is expected at long time, but rather a Kosterlitz-Thouless (KT) transition \cite{Kosterlitz1973}. For homogeneous quantum gases at equilibrium, the KT transition has been observed in liquid Helium \cite{Bishop1978} and with cold atoms \cite{Hadzibabic2006, Clade2009, Tung2010, Yefsah2011, Christodoulou2021}. In contrast, its spontaneous emergence in isolated systems following a quench is still  elusive, although progress  has been recently made in that direction \cite{Nazarenko2014, Situ2020}. Studies of the KT transition in the presence of disorder, on the other hand, remain scarce \cite{Plisson2011, Allard2012, Bourdel2012, Carleo2013}.

In this letter, we demonstrate 
the spontaneous emergence of a KT transition triggered by the sudden quench of a spatially disordered potential in a 2D interacting Bose gas, using
a relatively simple approach based on a random Gross-Pitaevskii equation. This
 allows us to explore two yet poorly understood problems: 
1- How 2D systems dynamically evolve toward the vicinity of a KT transition depending on the quench parameters and 
2- Once equilibrium is  established, how is the transition affected by the disorder. 
To address the first question, we numerically study the full time evolution of the coherence function and the thermalization time of the Bose gas. This allows us to explain, in particular, findings of the recent experiment \cite{Abuzarli2022}. On the second problem, we characterize the long-time, equilibrium transition line as a function of the disorder and compare it with state-of-the-art theoretical predictions. 
We finally investigate two core properties of the KT transition, the finite value of the superfluid density and the divergence of the correlation length, and find evidence for their universality in the presence of disorder.

Consider a dilute, 2D Bose gas initially prepared in a plane-wave state $|\bk_0\rangle$, i.e., an eigenstate of the free Hamiltonian. 
From now on we choose the initial
 momentum $\bk_0$ oriented along $x$. 
At time $t=0$, we assume that the gas is suddenly subjected to a 2D disorder potential $V(\boldsymbol{r}=x,y)$, which we model by an uncorrelated random Gaussian function with zero mean: $\overline{V(\boldsymbol{r})}=0$ and $\overline{V(\br)V(\br')}=\gamma\delta (\br-\br')$, where the overbar refers to the disorder average and $\gamma$ is the disorder strength. 
The latter defines an energy scale, $\gamma m$, where $m$ is the mass of the particles and we have set $\hbar=1$.
From $t=0$ onward, $|\bk_0\rangle$ is no longer an eigenstate of the problem, so that the Bose gas starts evolving in time. We describe this evolution using the nonlinear Schr\"odinger equation
\begin{equation}
i\partial_t\psi= -{\nabla}^2\psi/(2m) + V\psi+gN|\psi|^2
\psi,
\label{eq:grosspitaevskii}
\end{equation}
where $\psi=\psi(\br,t)$ is the wave function. 
The latter is normalized according to $\int d^2\br |\psi(\br,t)|^2=1$, and the pre-quench state is $\psi(\br,0)=1/\sqrt{\Omega}\exp(i\bk_0\cdot\br)$, with $\Omega$ the volume of the system. The particle density is $\rho_0=N/\Omega$. In practice, this quench protocol can be realized by cooling a Bose gas to low temperatures, transferring it a momentum and subjecting it to an optical random potential \cite{Jendrzejewski2012}. The problem is also relevant in the context of nonlinear paraxial optics where $\bk_0$ refers to the transverse wave vector of a laser impinging on a nonlinear medium at finite angle of incidence \cite{Fontaine2018, Santic2018, Sun2012}. In that case, the disorder can be realized by imprinting 2D refractive-index fluctuations in the $(x,y)$ plane \cite{Schwartz2007, Cherroret2018}.
\begin{figure}[h]
\centering
\includegraphics[scale=0.60]{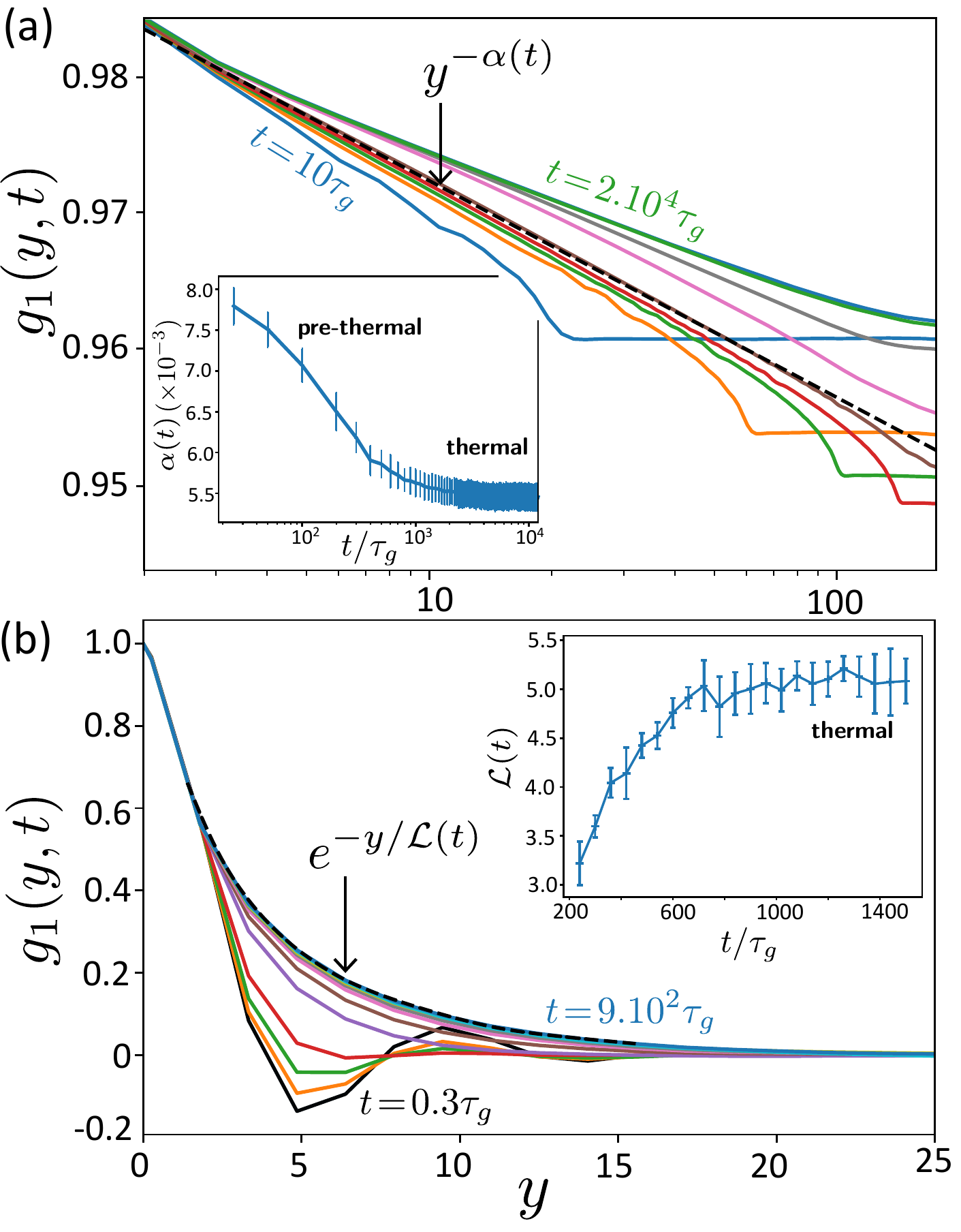}
\caption{
Coherence function  $g_1(x=0,y,t)$ vs. time of a 2D  Bose gas after a disorder quench, for two sets of initial conditions (the first and last time steps are indicated in each plot)
(a) $\gamma m=0.09$, $g\rho_0=1$ and $k_0=0$ ($\gamma m/g\rho_0\ll 1$):
shortly after the quench, $g_1$ decays algebraically, Eq. (\ref{eq_g1algebraic}).
At short time, the gas lies in a `pre-thermal' regime: the algebraic exponent $\alpha(t)\simeq \gamma m/4\pi g\rho_0$ (dashed line), then decreases slowly, and $g_1$ is limited by a Lieb-Robinson bound. At long time, thermal equilibrium establishes: $\alpha(t)\to\alpha(\infty)$ and $g_1$ is only limited by the system size. The inset shows the full evolution of $\alpha(t)$.
(b) $\gamma m=0.09$, $g\rho_0=0.003$ and $k_0=0.754$ ($\gamma m/g\rho_0\gg1$):
shortly after the quench, $g_1$ acquires an exponential shape, Eq. (\ref{eq_g1exp}), with a correlation length $\mathcal{L}(t)$ slowly increasing with time.
All data are averaged over $128$ disorder realizations and over system sizes $L\in[75,100]$ (a) and $L\in[350, 400]$ (b).
\label{fig_g1_time}}
\end{figure}

Averaging the solutions of the random Eq. (\ref{eq:grosspitaevskii}) over many realizations of $V(\br)$ allows us to effectively go beyond the mean-field level and capture the complex dynamics of the Bose gas, including the KT transition. 
In spirit, this approach is similar to a truncated Wigner approximation where one samples a random initial state \cite{Sinatra2002}.
To describe the gas dynamics, 
we study the average coherence function 
\begin{equation}
g_1(\br,t)={\overline{\psi^*(0,t)\psi(\br,t)}}/\overline{|\psi(\br,t)|^2}.
\label{eq:g1}
\end{equation}
To evaluate $g_1$, we numerical propagate the wave function with Eq. (\ref{eq:grosspitaevskii}) using a split-step algorithm, from which we compute the  momentum distribution \cite{Scoquart2020}. $g_1$ follows from inverse Fourier transformation and disorder averaging. In our simulations, we discretize space on a rectangular lattice $\Omega=L\times L$ with step $\delta=1.5$ and use periodic boundary conditions. Throughout the Letter, lengths, momenta and energies are given in units of $a$, $a^{-1}$ and $1/(ma^2)$, where $a$ is an arbitrary unit length.
The behavior of $g_1$ as a function of time is entirely governed by three independent energy scales characterizing the post-quench state: the kinetic, disorder and interaction energies, $k_0^2/2m$, $\gamma m$ and $g\rho_0$, respectively \cite{Footnote}. 
We first show in Fig. \ref{fig_g1_time} $g_1$ vs. time for two different sets $(k_0^2/2m,\gamma m,g\rho_0)$. 
In panel (a), 
$\gamma m\ll g\rho_0$.
In this low-energy regime (weak disorder quench), 
$g_1$ quickly exhibits an algebraic scaling:
\begin{equation}
\label{eq_g1algebraic}
g_1(\br,t)\sim \left({\xi}/{r}\right)^{\alpha(t)}.
\end{equation}
The algebraic exponent $\alpha(t)$ is shown in the inset of Fig. \ref{fig_g1_time}(a). It decreases in time and saturates when $t\sim 10^4\tau_g$, with $\tau_g=1/(g\rho_0)$, indicating that the system has reached its final equilibrium state where the gas behaves as a \textit{superfluid}. The time evolution of $\alpha(t)$ features a slow cross-over from a pre-thermal regime at short time, where $\alpha(t)\simeq \gamma m/4\pi g\rho_0$ \cite{Scoquart2020b}, to a finite-temperature superfluid state at long time, where $\alpha(t)\to\alpha(\infty)$ is related to the superfluid density, as will be investigated below. At short time, the algebraic decay is limited by the Lieb-Robinson bound $r=2 c_s t$,  where $c_s=\sqrt{g\rho_0/m}$ is the speed of sound, while at long time it is limited by the system size.

Fig. \ref{fig_g1_time}(b) shows  $g_1(\br,t)$ for a different initial condition, such that $\gamma m\gg g\rho_0$.
In this regime, the coherence function right after the quench is governed by disorder scattering and is non-universal 
\cite{Cherroret2021}. Very quickly though, particle interactions take over and induce an exponential coherence that persists over all times:
\begin{equation}
\label{eq_g1exp}
g_1(\br,t)\sim \exp[-r/\mathcal{L}(t)].
\end{equation}
The correlation length $\mathcal{L}(t)$ is shown in the inset of Fig. \ref{fig_g1_time}(b). It increases in time and saturates when $t\sim 10^3\tau_g$, the Bose gas now reaching a purely \textit{thermal} state.
While full thermalization takes some time to establish, it is remarkable that, in this regime of stronger disorder, the exponential coherence emerges shortly after the quench. 
This result explains the recent observations of \cite{Abuzarli2022}, where a transition from algebraic to exponential coherence was observed in a non-equilibrium fluid of light despite the system being probed at relatively short times.

This dynamical analysis shows that, after the quench, the Bose gas thermalizes into a phase that crucially depends on the initial conditions $(k_0^2/2m,\gamma m,g\rho_0)$. To characterize the final equilibrium state more exhaustively,  we have performed a systematic analysis of the coherence function vs. time for a large set of initial conditions. For each of those, we have determined whether $g_1(\br,t\to\infty)$  behaves exponentially or algebraically \cite{Supplemental}.
\begin{figure}[h]
\centering
\includegraphics[scale=0.75]{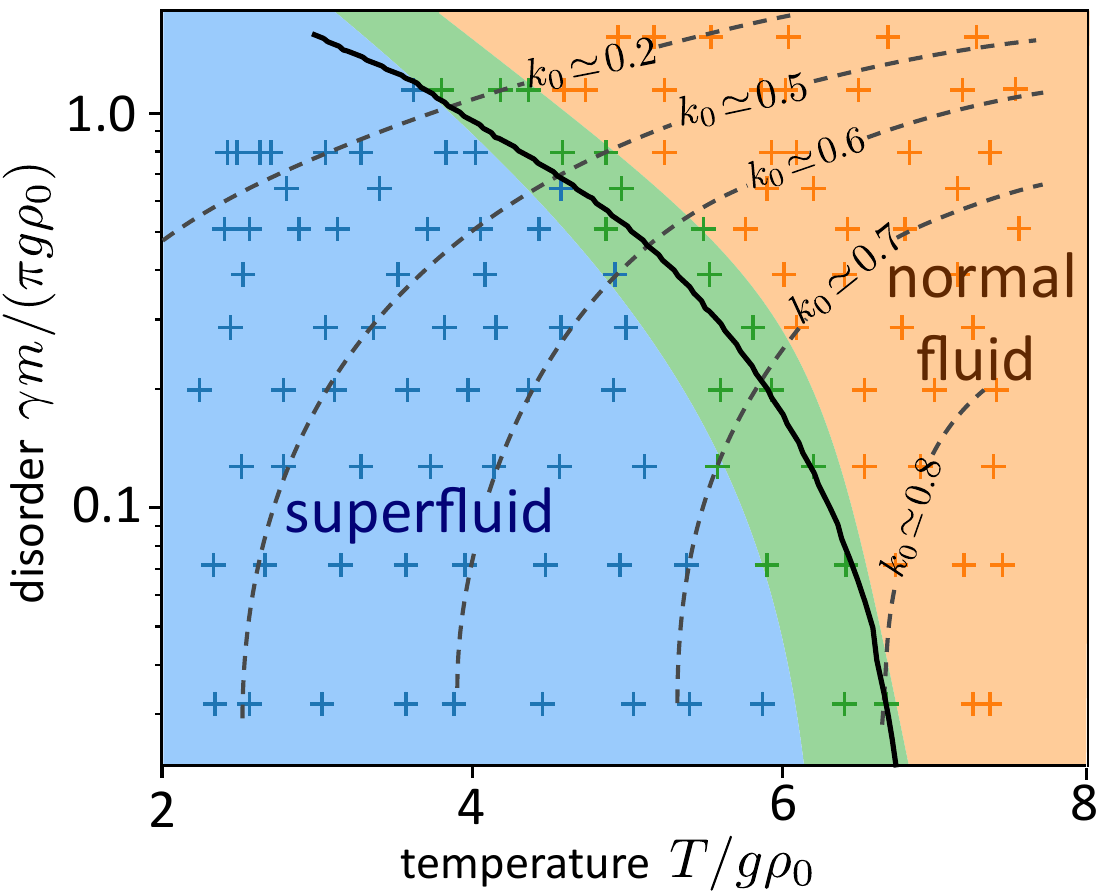}
\caption{
Equilibrium phase diagram reached by the 2D Bose gas a long time after the disorder quench, deduced from the spatial decay of $g_1(\br,t\to\infty)$. Each cross symbol corresponds to a given set $(k_0^2/2m,\gamma m,g\rho_0)$, with $\gamma m$ ranging from $0.01$ to $0.49$ and $k_0$ from 0 to $0.9$. Here $g\rho_0=0.1$ is fixed, and we use a system size $L=400$ and $64$ disorder realizations for averaging.
The dashed curves indicate the points of constant $k_0$ values, and the solid curve is the theoretical prediction for the Kosterlitz-Thouless transition line $(\gamma m, T_\text{KT})$, Eq. (\ref{eq_gammacritical}).
\label{fig_phasediagram}}
\end{figure}
The  result of this analysis is summarized by the phase diagram in Fig. \ref{fig_phasediagram}. The various equilibrium phases are represented as a function of the disorder energy $\gamma m$ and the temperature $T$, at fixed $g\rho_0$.  Here $T$ is the effective temperature acquired by the Bose gas after its self-equilibration, and is determined by the initial conditions.
To find it, we have computed numerically the occupation number at energy $\epsilon$, $N_\epsilon(t)=\rho_0\overline{\langle\psi(t)|\delta(\epsilon-H_0)|\psi(t)\rangle}/\nu_\epsilon$, where $H_0=-\nabla^2/(2m)+V$ and $\nu_\epsilon$  the average density of state per unit volume.
Whatever phase it lies in at long time, the Bose gas always contains a certain fraction of thermal particles. These particles occupy the states of highest energy, corresponding to $N_{\epsilon\to\infty}(\infty)\sim T/(\epsilon-\mu)$, where $\mu$ is the chemical potential.
This asymptotic law is the so-called Rayleigh-Jeans distribution, which describes the thermal equilibrium of classical-field theories \cite{Connaughton2005, Cherroret2015, Chiocchetta2016}. Therefore, by examining $N_\epsilon(\infty)$ at large $\epsilon$, one can infer both $T$ and $\mu$ for a given $(k_0^2/2m,\gamma m,g\rho_0)$ \cite{Supplemental}. In practice, at fixed $\gamma m$ and $g\rho_0$, larger temperatures are achieved by increasing $k_0$, as illustrated by the equi-$k_0$ lines in Fig. \ref{fig_phasediagram}. 

Fig. \ref{fig_phasediagram} shows that the long-time equilibrium state crosses from a superfluid to a normal fluid as $T$ is increased. A similar transition is also observed if the ratio $\gamma m/g\rho_0$ is increased at fixed temperature.
The set of parameters for which both an algebraic or exponential decay can equally well describe $g_1$ due to numerical uncertainties defines the central, green region in the phase diagram \cite{Supplemental}. It is in this region that we expect a Kosterlitz-Thouless transition to occur. 
To confirm it, we have evaluated semi-analytically $\gamma$ vs. the critical temperature $T_\text{KT}$
of the KT transition, in the spirit of the recent work \cite{Bertoli2018}. The approach consists in calculating the superfluid density $\rho_s(T)$ on the superfluid side of the transition  in the presence of disorder
using
 Bogoliubov theory \cite{Meng1994, Keeling2006}, and extrapolating the result to the transition point assuming 
the Nelson-Kosterlitz relation, $\rho_s(T_\text{KT})=2m T_\text{KT}/\pi$. 
This relation, well-known in the homogeneous case, describes  a universal jump  of the superfluid density at the transition \cite{Nelson1977}. 
Below we will verify its validity numerically in the presence of disorder. 
At weak disorder $\gamma m\ll g\rho_0$, this calculation provides
\begin{equation}
\label{eq_gammacritical}
\gamma m=\frac{4\pi g\rho_0}{I_1}\left[1-\frac{T_\text{KT}}{T_d}(4+I_2)\right],
\end{equation}
where $T_d=2\pi\rho_0/m$, $I_1=\int d\epsilon 4g\rho_0 (\nu_\epsilon/\nu)/(\epsilon+3g\rho_0-\mu)^2$ and $I_2=\int d\epsilon (\nu_\epsilon/\nu)/(\epsilon+3g\rho_0-\mu)$ with $\nu=m/2\pi$ \cite{Supplemental}. In these integrals, $\epsilon$ is bounded from above by the lattice energy $4/(m\delta^2)$. We also restrict ourselves to $\epsilon>0$, the density of states being very small at negative energy.
In the homogeneous limit ($\gamma, \delta\to0$), Eq. (\ref{eq_gammacritical}) reduces to $T_\text{KT}=T_d/\ln(e^4 T_d/g\rho_0)$, very close to the value $\simeq T_d/\ln(60 T_d/g\rho_0)$ obtained in Monte Carlo simulations \cite{Prokofev2001, Carleo2013}. Because Eq. (\ref{eq_gammacritical}) holds at weak disorder only, however, it is insufficient to  accurately capture our simulations. To solve this issue, we have further included the next-order disorder correction to Eq. (\ref{eq_gammacritical}) \cite{Yukalov2007b, Yukalov2007} and have also accounted for the disorder dependence of $\nu_\epsilon$ \cite{Supplemental}. With these corrections, Eq. (\ref{eq_gammacritical}) becomes an implicit equation for $\gamma$ which needs to be numerically solved. This yields the critical curve shown in Fig. \ref{fig_phasediagram}, where we have adjusted $g\rho_0\simeq 0.122$. This analysis confirms the validity of Eq. (\ref{eq_gammacritical}) within $20\%$ accuracy, a value that might be improved by working at weaker interaction.

To further characterize the KT transition emerging after the quench, we have studied two critical observables. The first one is the superfluid fraction $\rho_s(T=T_\text{KT})=2m T_\text{KT}/\pi$ at the transition, which in the absence of disorder is known to undergo a jump associated with a proliferation of vortices. This is also the relation we have assumed above for deriving the critical line, Eq. (\ref{eq_gammacritical}). 
\begin{figure}[h]
\centering
\includegraphics[scale=0.55]{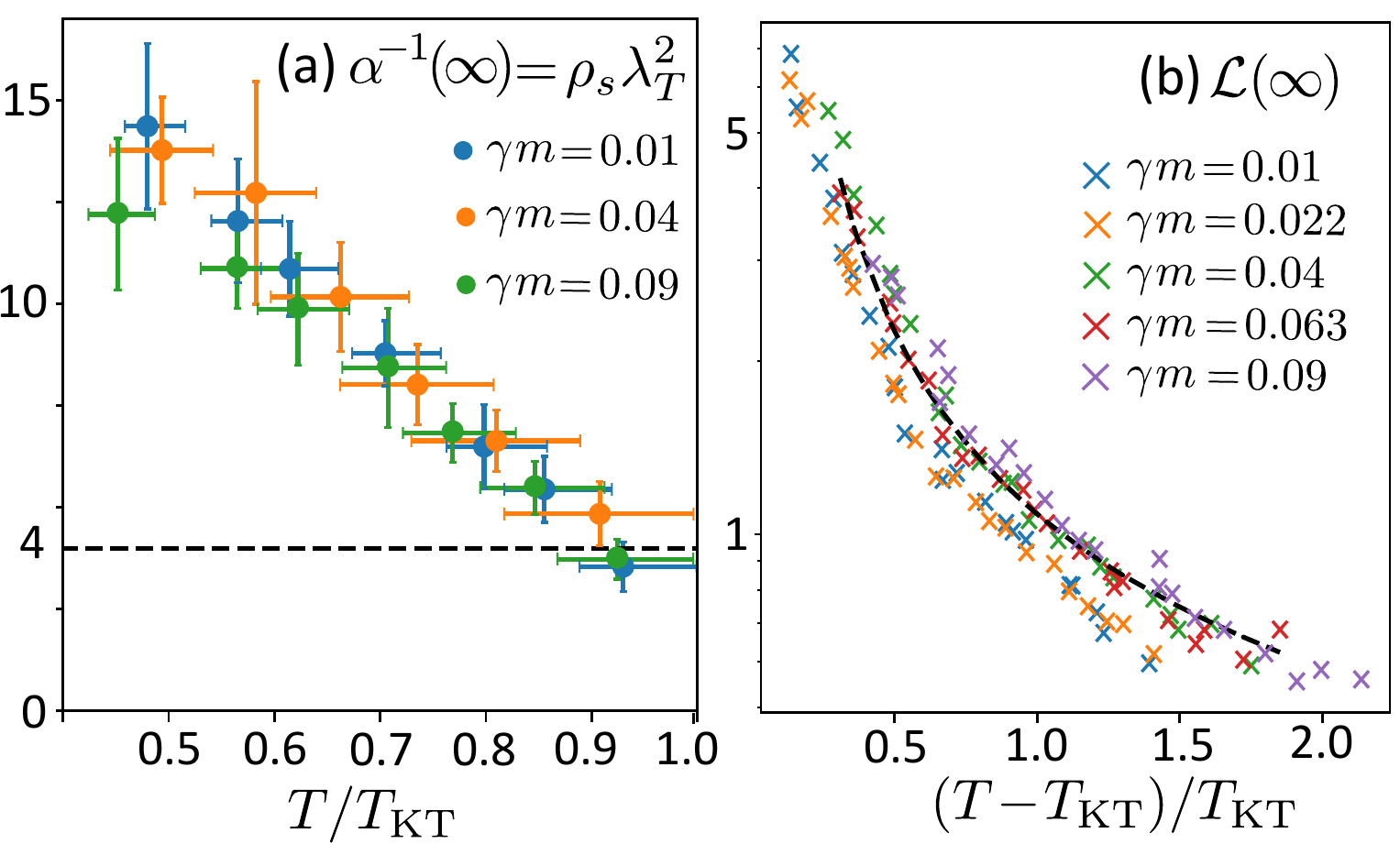}
\caption{
(a) Inverse of the algebraic exponent $\alpha(\infty)$ on the superfluid side of the KT transition, extracted from $g_1(\br,t\to\infty)$ for different temperatures and disorder strengths. The plot suggests $\alpha\simeq 1/4$ near the transition, irrespective of $\gamma$. The dashed line pinpoints $\alpha^{-1}(\infty)=4$.
(b) Correlation length $\mathcal{L}(\infty)$ extracted from $g_1(\br,t\to\infty)$ on the normal side, for different temperatures and disorder strengths. In both plots parameters are the same as in Fig. \ref{fig_phasediagram}, namely $g\rho_0=0.1$, $L=400$, and we average over $64$ disorder realizations. The critical temperature $T_\text{KT}$ is found numerically by taking the average of the two extreme points lying in the critical area in Fig. \ref{fig_phasediagram}.
The dashed curve is a fit to Eq.  (\ref{eq_LT_critical}).
\label{Critical_plots}}
\end{figure}
To test it, we have numerically computed the algebraic exponent $\alpha(t\to\infty)$ of $g_1(\br,t\to\infty)$, see Eq. (\ref{eq_g1algebraic}), for various disorder strengths and temperatures in the vicinity of $T_\text{KT}$. In homogeneous systems, $\alpha(\infty)=1/\rho_s(T)\lambda_T^2$, with $\lambda_T=\sqrt{2\pi/mT}$ the thermal wavelength. The jump of $\rho_s$ at the transition then corresponds to $\alpha(\infty)=1/4$. The results of this analysis are shown in Fig. \ref{Critical_plots}(a). As $T\to T_\text{KT}$, we indeed observe that points at different $\gamma$ tend to all satisfy $\alpha(\infty)=1/4$. 
A second central property of the KT transition is the fast divergence of the correlation length $\mathcal{L}$, see Eq. (\ref{eq_g1exp}), in the vicinity of the transition \cite{Hadzibabic2011}:
\begin{equation}
\label{eq_LT_critical}
\mathcal{L}(\infty)\sim\sqrt{\frac{T_\text{KT}}{T}}\exp\left[\sqrt{\frac{\zeta T_\text{KT}}{T-T_\text{KT}}}\right].
\end{equation}
To test Eq. (\ref{eq_LT_critical}) in the presence of disorder, we have extracted $\mathcal{L}(\infty)$ from $g_1(\br,t\to\infty)$ for various disorder strengths and temperature close to $T_\text{KT}$. The results, shown in Fig. \ref{Critical_plots}(b),
remarkably fall on the same universal curve. A fit (dashed curve) further demonstrates a good agreement with Eq. (\ref{eq_LT_critical}). Overall, these results point toward a universal character of the KT transition in the presence of disorder, once the critical temperature has been properly rescaled according to Eq. (\ref{eq_gammacritical}).

\begin{figure}[h]
\centering
\includegraphics[scale=0.32]{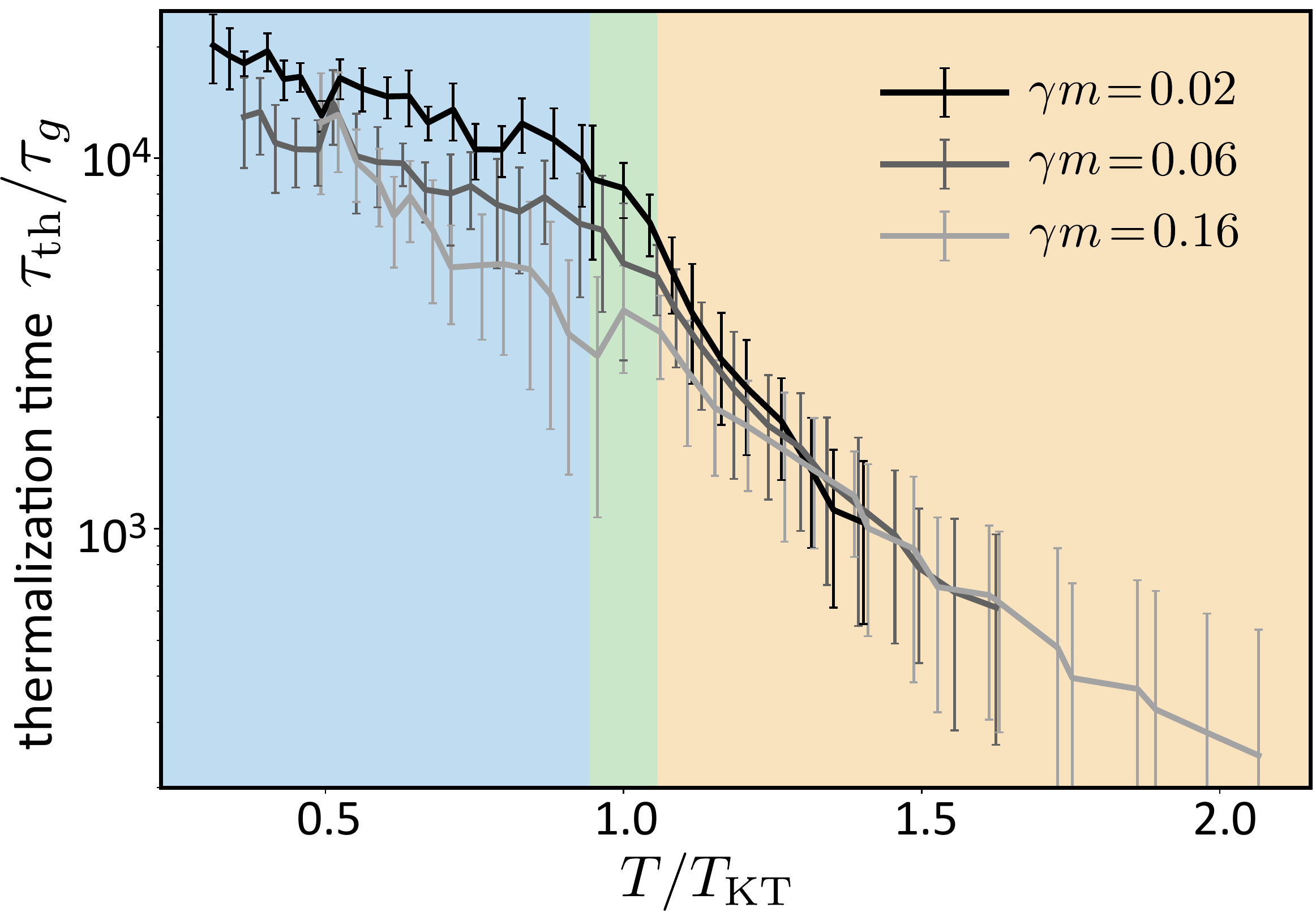}
\caption{
Thermalization time vs. temperature, for three disorder strengths. 
Parameters are the same as in Fig. \ref{fig_phasediagram}, except the system size $L=300$.
\label{Fig_tauth}}
\end{figure}
We have, finally, analyzed the thermalization time $\tau_\text{th}$ needed for the system to equilibrate after the quench. To estimate it, we use the empirical numerical criterion that beyond $\tau_\text{th}$, the area  $\smash{A(t)=\int_0^{L/2} dy \,g_1(x=0,y,t)}$ does not vary significantly in time whatever the quench parameters, i.e., for any point of the phase diagram in Fig. \ref{fig_phasediagram}. In practice, we fit $A(t)$ with a function of the form $B-C/(t+D)^2$, and define $\tau_\text{th}$ as the time for which this function reaches $95\%$ of its maximum $B$. The uncertainty on the fit parameters provide error bars for the determination of $\tau_\text{th}$. 
Fig.~\ref{Fig_tauth} shows $\tau_\text{th}$ vs.  temperature for different $\gamma m$. 
While $\tau_\text{th}(T)$ decreases nearly exponentially on the normal side, we observe a sizable increase of the thermalization time as one crosses from the normal to the superfluid phase. Notice also that, except perhaps at strong disorder, in the superfluid region disorder tends to `help' the system to thermalize faster. This nontrivial outcome is the result of a compromise between two antagonistic effects, the disorder that makes the system more ergodic after the quench, and the interactions that drive a superfluid state less sensitive to the disorder.

In conclusion, we have characterized the dynamical equilibration of 2D interacting Bose gases subjected a disorder potential quench, exploring both the non-equilibrium and equilibrium  facets of the problem and providing a benchmark for future experiments on atomic or photonic disordered fluids. An analysis of the dynamical exponents governing the approach to equilibrium would be an interesting challenge for future work.

NC acknowledges the Agence Nationale de la Recherche (grant ANR-19-CE30-0028-01 CONFOCAL) for financial support, and discussions with Andrea de Luca and Fabien Alet.

\clearpage
\section{ \Large{Supplemental Material}}

\section{Equilibrium temperature and chemical potential}

\subsection{Energy distribution}
\label{energy_distrib_sec}

To numerically determine the  temperature $T$ and the chemical potential $\mu$ of the thermal equilibrium for a given set $(k_0^2/2m, \gamma m,g\rho_0)$ of initial quench parameters, we consider the disorder-averaged occupation number at long time, defined as
\begin{equation}
 N_\epsilon=\lim_{t\to\infty}\frac{{\rho_0}}{\nu_\epsilon}\overline{\langle\psi(t)|\delta(\epsilon-H_0+g|\psi(t)|^2/2)|\psi(t)\rangle}.
 \label{feps_diff}
 \end{equation}
Here $\rho_0$ is the particle density,  $\nu_\epsilon$ is the average density of states per unit volume, $H_0=-\nabla^2/(2m)+V$, and $|\psi(t)\rangle$ is the wavefunction of the Bose gas at time $t$. As in the main text, we set $\hbar=1$. In the following, every dimensional quantity is expressed through a chosen unit of length $a$: the density $\rho_0$ is in unit of $1/a^2,$ wavevectors are in units of $1/a$, and energies and temperatures in units of $1/(ma^2)$. 
The  conservation of particles and energy impose $\int{ d}\epsilon\,\nu_\epsilon {N_\epsilon} = {\rho_0}$ and $\int{ d}\epsilon\,\nu_\epsilon\epsilon {N_\epsilon}={\rho_0}[{k_0^2}/(2m)+{g\rho_0}/{2}]$.

Since the particles evolve in a disordered potential, here chosen of Gaussian statistics, the energies $\epsilon$ are not bounded from below and, in particular, can be negative. At low energy, the behavior of $N_\epsilon$ is \textit{a priori} complicated. It turns out, however, that in order to access the temperature and the chemical potential in the equilibrium state, it is sufficient to consider the asymptotic expression of $N_\epsilon$ at large energy, which we now discuss. This asymptotic limit takes a different form on both sides of the Kosterlitz-Thouless (KT) transition. 
At high enough temperatures first (i.e., above the KT transition), the gas is mainly composed  of thermal particles. Within the classical-field description considered in our work and for weak enough disorder, its high-energy asymptotic limit is well described by the Rayleigh-Jeans distribution \cite{Connaughton2005, Cherroret2015, Chiocchetta2016}
\begin{equation}
N_\epsilon\simeq \frac{T}{\epsilon-\mu}.
\label{eq:rj_distrib}
\end{equation}
At lower temperatures (i.e., below the KT transition) on the other hand, a fraction of the gas becomes superfluid and coexists with the thermal fraction of atoms. The superfluid fraction is associated with the presence of Bogoliubov quasi-particles. In the limit of weak disorder $\gamma m\ll g\rho_0$, the average occupation number $N^\text{qp}_\epsilon$ of the quasi-particles is given by~\cite{Yukalov2007b}:
\begin{equation}
N^\text{qp}_{\omega_\epsilon}
=\frac{T}{\omega_\epsilon},
\label{eq:f_eps_bogo}
\end{equation}
where $\omega_\epsilon=\sqrt{(\epsilon+3g\rho_0-\mu)(\epsilon+g\rho_0-\mu)}$ is the Bogoliubov dispersion relation in a non-uniform system, and $\mu$ is the chemical potential associated with the normal fraction of the gas. Note that while $\mu=g\rho_0$ in the homogeneous limit, its value for a discrete disordered system is a priori not known. From Eq. (\ref{eq:f_eps_bogo}), the occupation number of the true particles follows from $N_\epsilon d\epsilon=N^\text{qp}_{\omega_\epsilon}d\omega_\epsilon$. At high energy $\epsilon\gg g\rho_0$, this gives:
\begin{equation}
{N_\epsilon}\simeq\frac{T}{\epsilon+4g\rho_0-2\mu}.
\label{f_eps_bogo_2}
\end{equation}
Thus, on either side of the transition the high-energy limit of the average occupation number is of the form $T/(\epsilon-C)$, where $C$ is related to the chemical potential. We have used this simple property to numerically estimate $T$ and $\mu$ for any value of the quench parameters, as we now explain.

\subsection{Numerical evaluation of ${N_\epsilon}$}

To obtain an estimation of ${N_\epsilon}$, we numerically compute the temporal evolution of ${N_\epsilon}(t)$ and consider that the equilibrium value is reached when ${N_\epsilon}(t)$ no longer depends on time. For convenience, in our numerical experiments we choose $g\rho_0$ small enough so that, in a first approximation, we can omit the interaction term when evaluating the energy distribution, Eq.~(\ref{feps_diff}), and use: 
\begin{equation}
 \nu_\epsilon {N_\epsilon}(t)\simeq{\rho_0}\overline{\langle\psi(t)|\delta(\epsilon-H_0)|\psi(t)\rangle},
 \label{feps_diff2}
 \end{equation}
where $\psi(t)$ is the wavefunction of the Bose gas after an evolution time $t$.
To evaluate $\nu_\epsilon{N_\epsilon}(t)$ numerically, we express it as the Fourier transform of the time-autocorrelation function of the gas~\cite{Trappe2015}, defined by $\mathcal{C}_t(\tau) = \overline{\langle\psi(t)|\exp(-iH_0\tau)|\psi(t)\rangle}$:
\begin{equation}
  \nu_\epsilon {N_\epsilon}(t)= \frac{{\rho_0}}{\pi} \Re \int_0^\infty d\tau \,\,\mathcal{C}_t(\tau) \ e^{i\epsilon\tau}.
\label{eq:fourier_autocorr}
\end{equation}
We first obtain $\psi(t)$ by propagating the initial plane wave $\ket{\bk_0}$ for a time $t$ with the Gross-Pitaevskii equation. Then $\psi(t)$ is used as the initial state for another propagation with $H_0$ for a time $\tau$, which yields the desired $\mathcal{C}_t(\tau)$. We finally perform numerically the Fourier transform (\ref{eq:fourier_autocorr}) and average over disorder realizations. 

To access ${N_\epsilon}$, the last step consists in computing 
the density of states per unit volume of the disordered system, $\nu_\epsilon=\int{{ d}^2 \bk}/{(2\pi)^2}\overline{\langle\bk|\delta(\epsilon-H_0)|\bk\rangle}$. We achieve this task separately by exploiting the statistical translation invariance, which implies $\nu_\epsilon=\overline{\langle\br|\delta(\epsilon-H_0)|\br\rangle}$ for any  point $\br$. We  then use the same procedure as explained above, with the time-autocorrelation function $\mathcal{C}(\tau)=\overline{\langle\br |e^{-iH_0\tau}|\br\rangle}$ now computed by propagating the point-like state $|\br\rangle=|0\rangle$ with $H_0$.

\begin{figure}
	\centering
	\includegraphics[width=0.9\linewidth]{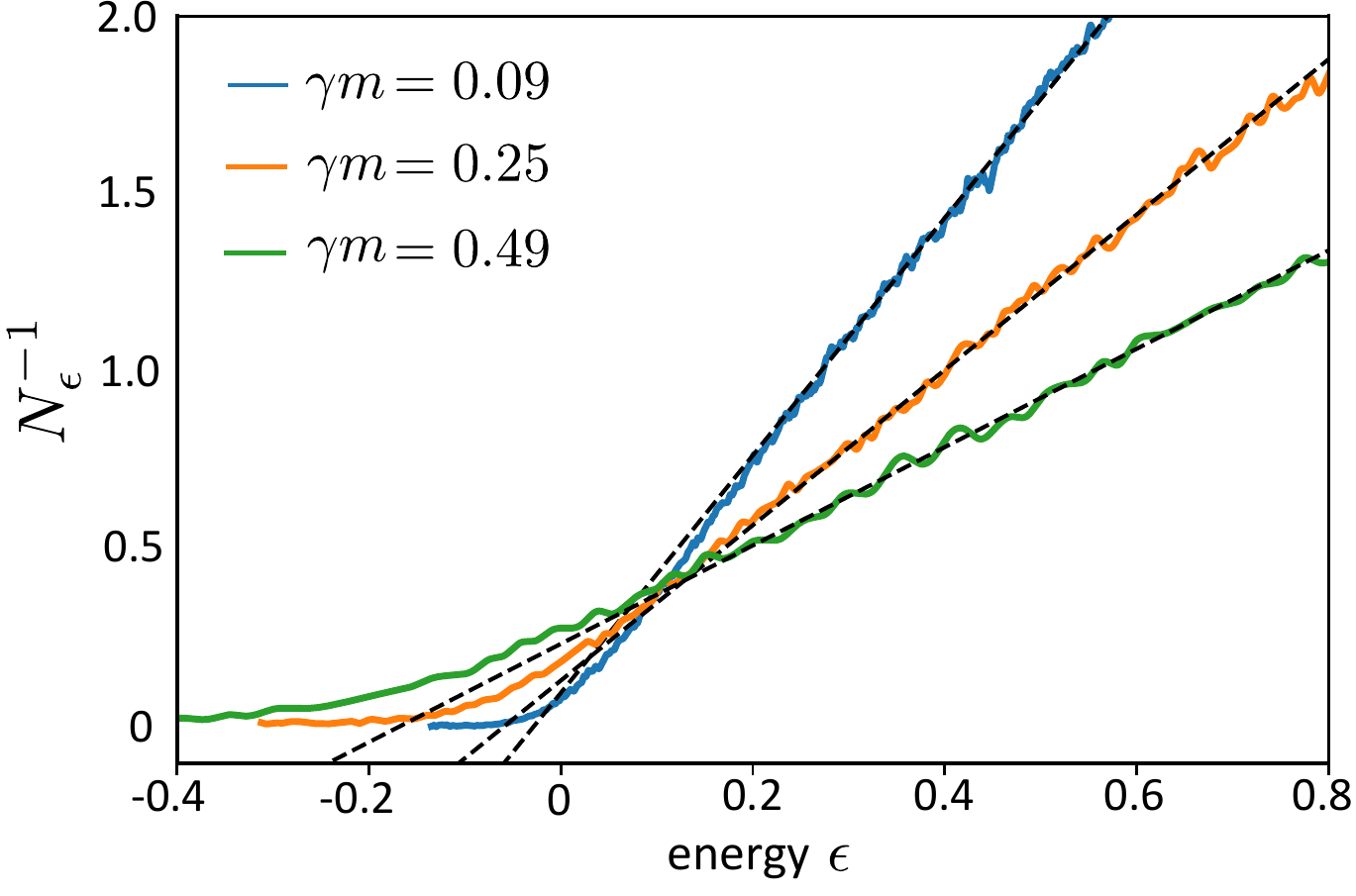}%
	\caption{Examples of simulated inverse occupation numbers ${N_\epsilon^{-1}}$ at thermal equilibrium, with fixed quench parameters $k_0=0.45$ and $g\rho_0=0.1$ and for three values of the disorder energy $\gamma m$. At high energy, the data are perfectly fitted by $y=(x-C_1)/T$ (dashed lines). In the normal region of the equilibrium phase diagram, $C_1=\mu$ according to Eq. (\ref{eq:rj_distrib}). In the superfluid region, $C_2=-4g\rho_0+2\mu$, see Eq. (\ref{f_eps_bogo_2}).
We recall that the units of momentum and energy are $a^{-1}$ and $(ma^2)^{-1}$, respectively (with $a$ the chosen unit of length).
		\label{fig:1_over_feps}}
\end{figure}

Once $\nu_\epsilon {N_\epsilon}$ and $\nu_\epsilon$ computed, we plot the ratio ${N_\epsilon}(t)^{-1}=\nu_\epsilon\times (\nu_\epsilon {N_\epsilon})^{-1}$. As explained in Sec. \ref{energy_distrib_sec}, at large enough energy this quantity is expected to scale linearly with $\epsilon$, see Eqs. (\ref{eq:rj_distrib}) and (\ref{eq:f_eps_bogo}). This is indeed observed numerically, as illustrated in Fig.~\ref{fig:1_over_feps} for three different values of the disorder energy $\gamma m$. From these plots, we access $T$ and $\mu$ from a simple linear fit.

Note that, in practice, this way of extracting the final temperature and chemical potential sets a lower limit for the value of the disorder parameter $\gamma$ providing reliable results. Indeed, when the mean free path of a given diffusive mode $\ell_k\sim k/(\gamma m^2)$ starts to exceed the system size, the partial wave amplitudes propagating with momentum $k$ in the medium -- which typically decay as $\exp(-r/\ell_k)$ -- become non-negligible as they reach the edges of the system, which causes them to self-interfere because of the boundary conditions. This makes the temporal autocorrelation of the wavefunction increase by a constant amount, which translates into parasitic oscillations in the energy distribution. This problem is all the more present when probing long times, as the momentum distribution spreads out, and modes of high momenta associated with a large mean free path are populated. A good condition for this effect to be negligible is that the mean free path associated with the maximum possible value of the momentum is smaller than $L/10$. As we discretize spatially the system over a grid of step $\delta,$  the maximum momentum is $k=\pi/\delta,$ the associated mean free path $\ell_\text{max}=\pi/(\delta \gamma m^2),$ so that the condition reads $\gamma \geq 10\pi/(\delta L m^2)$. Note that whenever this condition is fulfilled, we checked that the computed temperature and chemical potential have a very weak dependence on the size of the system, indicating  that the system is very close to the true thermodynamic limit.

\section{Construction of the phase diagram}

To build the long-time equilibrium phase diagram of the Bose gas (Fig. 2 of the main text), 
we have numerically performed extensive temporal propagations of the initial state $|\bk_0\rangle$ for a wide range of initial parameters $(k_0^2/2m, \gamma m)$, at fixed 
interaction strength $g\rho_0=0.1$, spatial discretization step $\delta=1.5$ and system size $L=400\simeq 126\xi$, where $\xi=1/\sqrt{mg\rho_0}$ is the healing length. 
We have then computed both the $g_1$ coherence function and the energy distribution ${N_\epsilon}$ of the Bose gas, up to a time of $t_\text{max}=10^4\tau_g$ with $\tau_g=(g\rho_0)^{-1}$. At such long times, the coherence function becomes independent of time for most choices of parameters.

Once a temporal propagation is completed, we first extract the temperature $T$ from the energy distribution, as explained in the previous section. Using this value, each set of initial parameters maps onto a single point in the $\gamma m-T$ phase diagram of the main text.

To characterize the equilibrium phase of the gas reached after the evolution, we use the following empirical procedure. Starting from the somewhat arbitrary length $3\xi$, we try to fit a cut of the $g_1(\br,t\to\infty)$ along the $y$-axis with positive $y$ (i.e., in the direction perpendicular to the initial momentum $\bk_0$) by an algebraic law, up to the length $25\xi$ to avoid any numerical noise stemming from the use of periodic boundary conditions. 
If the algebraic fit is successful, i.e., with an estimated relative error on the fit parameters less than $1\%$, we conclude that the gas lies in the superfluid phase, indicated in blue in the phase diagram. If the algebraic fit is not successful, we try to perform an exponential fit from $3\xi$ up to the point $y$ where $g_1<10^{-2}$. A successful fit then indicates that the gas lies in the normal phase, indicated in orange in the phase diagram. 

\begin{figure}
	\centering
	\includegraphics[width=0.95\linewidth]{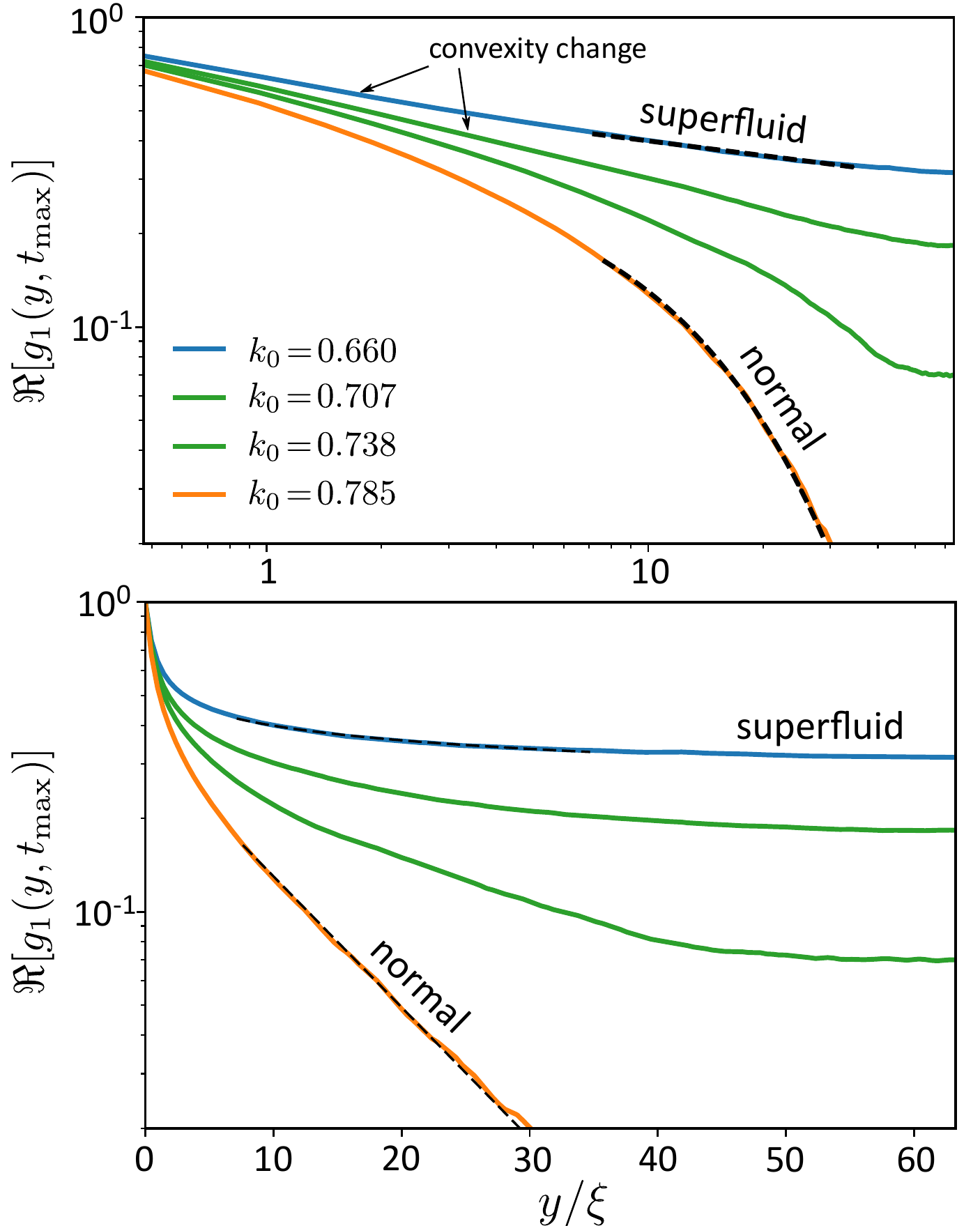}%
	\caption{
Examples of time-independent $g_1$ functions after a propagation time $t_\text{max}$, in the three different phases of the phase diagram, for $g\rho_0=0.1$, at fixed disorder strength $\gamma\simeq 0.06$ for three values of $k_0$ indicated in the upper plot. Here $t_\text{max}=10^{4}\tau_g$, sufficiently large so that $\Re[g_1(t_\text{max})]$ is stationary for these parameters.
Depending on $k_0$, the gas lies either in the normal (orange curve), superfluid (blue curve) or transition (green curves) region at long time (the dashed curves correspond to an exponential or an algebraic fit, respectively).
The upper plot is on a doubly logarithmic scale, the lower one on a semi-logarithmic scale. 
The log-log plot of $g_1$ at short scale 
changes from a convex to a concave shape when crossing from the superfluid to the normal side.
		\label{fig:g1_examples}}
\end{figure}

In the vicinity of the KT transition, the fitting procedure becomes more challenging. Indeed, as shown in Fig.~\ref{fig:g1_examples}, close to the transition the $g_1$ function is neither completely algebraic nor exponential. 
This defines a zone (represented in green), where both fits may work approximately. On the normal side, we then simply consider that the gas is in a normal-fluid state when the exponential fit is nearly perfect, for values of $g_1$ above $10^{-2}$. This criterion yields a precise distinction between the two phases because the characteristic decay length of the exponential diverges very sharply as we approach the critical temperature \cite{Kosterlitz1973,Hadzibabic2011}. 
On the other side of the green area, where $g_1$ decays rather slowly, things are less clear and relying on the accuracy of the algebraic fits yields a blurry separation between the blue and the green zones. However, we have noticed that the log-log plot of the coherence function at short scale suddenly changes from concave to convex as one approaches the KT transition from the superfluid side (see Fig.~\ref{fig:g1_examples}). We take this change of convexity as our criterion for the distinction between the blue and the green zones.  This empirical procedure yields a rather wide critical line, which is expected to contain the critical temperature of the KT transition. This statement is confirmed by a good agreement with the theoretical prediction of the critical temperature (or critical disorder), represented by the solid curve in Fig. 2 of the main text. 
Finally, we have checked that the general shape of the phase diagram is unaffected by small variations of the value of the interaction strength or by slight changes of the discretization step.

\section{Critical disorder strength}

In this section, we recall the main theoretical steps leading to the critical disorder for the KT transition, Eq. (5) of the main text. To this aim, we first follow the reasoning developed in \cite{Bertoli2018} and \cite{Yukalov2007b}, and then adapt it to our discrete, two-dimensional system.
As in \cite{Bertoli2018}, here we restrict ourselves to small interactions, $g\rho_0\ll T$, a condition that is essentially verified in our simulations in the vicinity of the KT transition.

\subsection{Weak disorder}

In the presence of a weak disorder potential such that $\gamma m\ll g\rho_0$, the expression for the normal-fluid density $\rho_n$ of a two-dimensional, weakly interacting Bose gas, was previously derived in \cite{Meng1994} within the Bogoliubov framework. It consists in computing the static linear response function of the system, which naturally separates into a normal and a superfluid part \cite{Keeling2006}. For our discrete system at leading order in $\gamma m$ and $g\rho_0$, the normal-fluid density reads: 
\begin{align}
\rho_{n}&=\frac{\rho_0}{2}\int\frac{{\rm d^2}\bk}{(2\pi)^2}\frac{\tilde{\mathcal{C}}(\bk)}{(g\rho_0+\epsilon_\bk/2)^2}-\int \frac{{\rm d^2}\bk}{(2\pi)^2} \epsilon_\bk\frac{\partial {N^\text{qp}_{\omega_\bk}}}{\partial \omega_{\bk}}\\&\equiv\rho_{n}^\text{dis}+\rho_{n}^\text{hom},
\label{eq:normal_density_def_momentum}
\end{align}
where $\epsilon_\bk=E_\bk+g\rho_0-\mu$ with $E_\bk$ the kinetic energy of the particles, $\omega_\bk=\sqrt{(E_\bk+3g\rho_0-\mu)(E_\bk+g\rho_0-\mu)}$ is the Bogoliubov dispersion relation, $f^*_{\omega_\bk}$ is the occupation number (\ref{eq:f_eps_bogo}) of the quasiparticles, and $\tilde{\mathcal{C}}(\bk)=\gamma$ is the power spectrum of the disorder. For a discrete system, the kinetic energy is
\begin{equation}
E_{k_x,k_y}=\frac{2-\cos{(k_x \delta)}-\cos{(k_y \delta)}}{m\delta^2},
\label{eq:disp_rel_discr_chap5}
\end{equation}
with $\delta$ the discrete spatial step.

In the second equality of Eq. (\ref{eq:normal_density_def_momentum}), we introduced $\rho_{n}^\text{dis}$, the correction of the normal fluid density due to the disorder potential. The contribution $\rho_{n}^\text{hom}$, on the other hand, defines the normal-fluid density in the homogeneous (i.e., non-disordered) limit.
For weak interactions $g\rho_0\gg T $, the two contributions can be expressed as : 
\begin{align}
&\rho_{n}^\text{dis}=\frac{m\rho_0\gamma}{4\pi g\rho_0}I_1(g\rho_0,\gamma,\delta,\mu)
\label{eq:rho_dis},\\
&\rho_{n}^\text{hom}=\frac{mT}{2\pi}I_2(g\rho_0,\gamma,\delta,\mu),
\label{eq:rho_hom}
\end{align}
where 
\begin{align}
&I_1(g\rho_0,\gamma,\delta,\mu)=\int{\rm d}\epsilon\,\frac{4g\rho_0\nu_\epsilon/\nu}{(\epsilon+3g\rho_0-\mu)^2} ,
\label{eq:I1}\\
&I_2(g\rho_0,\gamma,\delta,\mu)=\int{\rm d}\epsilon\, \frac{\nu_\epsilon/\nu}{\epsilon+3g\rho_0-\mu}.
\label{eq:I2}
\end{align}
The two dimensionless constants $I_1$ and $I_2$ depend on the interaction strength $g\rho_0$, the spatial discretization $\delta$ and the chemical potential $\mu$. Their dependence on the disorder strength $\gamma$, due to disorder corrections to the density of states, is negligible at weak disorder. The quantity $\nu=m/(2\pi)$ is the free-space density of state of the \textit{continuous} problem. 
In Eqs. (\ref{eq:I1}) and (\ref{eq:I2}), the integrals run only over positive energies as we neglect the occupation of the low-energy states induced by the presence of disorder. Indeed, for sufficiently small disorder, the density of states $\nu_\epsilon$ remains small in the negative energy region. The upper bound is set to $4/m\delta^2$, the maximum of the dispersion relation (\ref{eq:disp_rel_discr_chap5}). Again, for small disorder, we can safely neglect the effect of the non-zero density of states in the high energy region.
Using the simulated values of $T$, $\mu$ and of the density of states $\nu_\epsilon$, we can numerically compute $I_1$ and $I_2$, which yields the normal-fluid density (\ref{eq:normal_density_def_momentum}) of the Bose gas at equilibrium :
\begin{equation}
\rho_{n}=\rho_{n}^\text{dis}+\rho_{n}^\text{hom}=\frac{m\rho_0\gamma}{4\pi g\rho_0}I_1+\frac{mT}{2\pi}I_2.
\label{eq:normal_density_I1I2}
\end{equation}
To obtain an expression for the critical disorder of the KT transition, we assume that the Nelson-Kosterlitz relation for the universal jump of the superfluid density at the transition is still valid in the presence of disorder~\cite{Nelson1977}, as one approaches the transition from the superfluid side:
\begin{equation}
\rho_s(T)= \frac{2m}{\pi}T\quad \text{at}\quad T= T_\text{KT}.
\label{eq:Nelson-Kosterlitz}
\end{equation} 
As explained in the main text, the behavior of the $g_1$ function on both sides on the KT transition fully supports this assumption.
By decomposing the gas density into a normal and a superfluid part, we infer, at $T= T_\text{KT}$,
\begin{equation}
\rho_n=\rho_0-\rho_s=\rho_0\left(1-4\frac{T_\text{KT}}{T_d}\right),
\label{eq:rho_n_nelson}
\end{equation}
where $T_d={2\pi\rho_0}/{m}$. From Eqs.~(\ref{eq:normal_density_I1I2}) and (\ref{eq:rho_n_nelson}), we obtain the following expression for the critical disorder strength $\gamma_\text{KT}$ corresponding to the Kosterlitz-Thouless transition:
\begin{equation}
\gamma_\text{KT}=\frac{4\pi g\rho_0}{m I_1}\left[1-\frac{T_\text{KT}}{T_d}\left(4+I_2\right)\right].
\label{eq:critical_disorder_1}
\end{equation}
At weak disorder, $I_1$ and $I_2$ depend very weakly on the disorder strength $\gamma$, and are easily computed from Eqs. (\ref{eq:I1}) and (\ref{eq:I2}) at a given interaction strength $g\rho_0$,  using the simulated density of states and the numerical values of $\mu$. 
Then, we can solve implicitly Eq.~(\ref{eq:critical_disorder_1}) for pairs of solutions $(\gamma_\text{KT},T_\text{KT})$.
Note that in the continuous limit, Eq. (\ref{eq:critical_disorder_1}) yields the same expression as in Ref.~\cite{Bertoli2018}:
\begin{align}
\epsilon^*(T_\text{KT})=2g\rho_0\left[1-\frac{T_\text{KT}}{T_d}\ln{\left(e^4\frac{T_\text{KT}}{g\rho_0}\right)}\right],
\label{eq:critical_disorder_ref}
\end{align}
with $\epsilon^*=\gamma_\text{KT}m/\pi$. In the absence of disorder, $\gamma_\text{KT}=0$ by definition, Eq. (\ref{eq:critical_disorder_ref}) gives for the critical temperature $T_\text{KT}\simeq T_d/\ln{(e^4T_d/g\rho_0)}$, which is very close to the value obtained by Monte-Carlo simulations in \cite{Prokofev2001}: $T_\text{KT}\simeq T_d/\ln{(\xi T_d/g\rho_0)}$, with $\xi\simeq 60$. According to Ref.~\cite{Bertoli2018}, in the regime of weak disorder $\gamma m\ll g\rho_0$, Eq.~(\ref{eq:critical_disorder_ref}) yields values of the critical temperature which agree within $20\%$ with the results of the Monte-Carlo numerical approach developed in \cite{Carleo2013}. \\

\subsection{Stronger disorder}

As the disorder becomes stronger, $\gamma m\sim g\rho_0$, $I_1$ and $I_2$ start to significantly depend on the disorder strength through a modification of the density of states $\nu_\epsilon$. As a result, at the KT transition Eq. (\ref{eq:critical_disorder_1}) becomes an implicit function of the critical disorder strength $\gamma_\text{KT}$. Another issue is the validity of our initial formula (\ref{eq:normal_density_def_momentum}), which was derived within a Bogoliubov approach assuming weak disorder. As $\gamma m$ approaches $g\rho_0$, this approach must be corrected by taking into account higher-order interaction terms between quasi-particles and the random potential in the Bogoliubov Hamiltonian. In three dimensions, such a task was achieved in \cite{Yukalov2007b}. 
We will not present here a full adaptation of the reasoning to the two-dimensional case, and refer the interested reader to the aforementioned work. 
The net impact of the next-order disorder corrections on formula (\ref{eq:critical_disorder_1}) is a renormalization of the interaction strength $g\rho_0$ via
\begin{align}
g\rho_0\to g\rho_0\left[1+\frac{\gamma}{2}\int{\rm d}\epsilon\,\frac{4\nu_\epsilon}{(\epsilon+3g\rho_0-\mu)^2}
\right].\label{eq:g_renorm}
\end{align}
The correction term in Eq. (\ref{eq:g_renorm}) turns out to coincide with the disorder contribution to the normal density in Eq. (\ref{eq:normal_density_def_momentum}). Computing it with our numerical values of the temperature, chemical potential and density of states, and performing the replacement (\ref{eq:g_renorm}) into Eq.~(\ref{eq:critical_disorder_1}), we obtain a new implicit equation for the critical disorder $\gamma_{KT}$, which we use to draw the critical solid curve in the phase diagram of Fig. 2 of the main text. 
It turns out, however,  that in our simulations the value of $g\rho_0$ is such that $T_\text{KT}/g\rho_0\sim 6$, which is not exactly within the limit $g\rho_0\ll T$ assumed above. In addition to the disorder corrections, we thus expect higher-order contributions in $g\rho_0$ to play a role. We empirically account for this effect by computing $I_1$ and $I_2$ using the interaction energy as a fit parameter $(g\rho_0)_\text{fit}=1.22g\rho_0$. The resulting transition line in reasonable agreement with the position of the KT transition observed in the numerical simulations. 


\begin{thebibliography}{99}
\bibitem{Footnote}
The lattice energy, $1/(m\delta^2)$, amounts to a global scaling factor and is thus fixed throughout the Letter.
\end{thebibliography}
\end{document}